\documentclass{jpsj-suppl}
\usepackage{times}
\usepackage{color}
\usepackage{ulem}
\title{
On the Stability of Quantum Hall Kagome-ice Insulator 
}

\author{
Hiroaki Ishizuka\thanks{E-mail address: ishizuka@aion.t.u-tokyo.ac.jp} and Yukitoshi Motome
}
\inst{
Dept. of Applied Physics, the University of Tokyo, 7-3-1 Hongo, Bunkyo, Tokyo 113-8656
}
\abst{
The stability of the kagome-ice insulator against the three dimensionality and the canting angle of the spin anisotropy axes is studied numerically using the exact diagonalization method.
The kagome-ice insulator, which was recently found at 2/3 electron filling in the double-exchange model on a kagome lattice, is a Chern insulator showing a quantized Hall response despite the absence of magnetic long-range order.
We find that this peculiar state remains stable when the lattice is extended to three dimensions in an anisotropic pyrochlore lattice in the weak interlayer coupling region. 
On the other hand, we show that the kagome-ice insulator is also stable against the change of the canting angle. 
Furthermore, we find another kagome-ice insulator at 1/3 filling in a different range of the canting angle. 
Our results indicate that the quantum anomalous Hall response is widely observed in the variants of the kagome-ice double-exchange systems.
}

\kword{
Kagome lattice, Kagome ice, Chern insulator, Quantum anomalous Hall effect, Metal-insulator transition
}

\begin{document}
\maketitle

\section{Introduction}

Competing interactions in geometrically frustrated magnets often give rise to the development of a peculiar local spin correlation at low temperatures.
A representative example is the two-in two-out spin configuration in each tetrahedron in pyrochlore spin ice~\cite{Harris1997,Ramirez1999}. 
In the spin ice, the ferromagnetic nearest-neighbor (NN) interaction between the Ising-like moments with the local $[111]$ anisotropy gives rise to the two-in two-out local correlation, which is called the ice rule, and the geometry of the pyrochlore lattice leads to the suppression of the long-range ordering.
Such correlated liquid-like states have been a matter of intense studies as they are the source of interesting properties in frustrated magnets, such as macroscopic degeneracy with residual entropy and a characteristic power-law spin correlation~\cite{Isakov2004,Henley2005}.

When such locally correlated spins are coupled to itinerant electrons, the scattering of the itinerant electrons by the correlated localized moments strongly affects the electronic state.
For instance, extended Falicov-Kimball models on frustrated lattices show non-Fermi liquid behavior~\cite{Udagawa2010} and peculiar metal-insulator transition~\cite{Ishizuka2011}.
A related study was done on a frustrated double-exchange (DE) model, focusing on the nature of loops emergent from local spin textures~\cite{Jaubert2012}. 
The scattering by correlated spin textures also gives rise to peculiar transport phenomena as well.
It was recently reported that the development of the local correlation in spin-ice type models gives rise to a resistivity minimum~\cite{Udagawa2012,Chern2013} and peculiar magnetic field dependence of the anomalous Hall effect~\cite{Udagawa2013}. 
The results were discussed in relation to the transport properties observed in Pr$_2$Ir$_2$O$_7$~\cite{Nakatsuji2006,Machida2007,Balicas2011}.
Thus, such spin-charge interplay offers a fertile ground for exploring unconventional electronic and transport properties.

\begin{figure}
\begin{center}
\includegraphics[width=\linewidth]{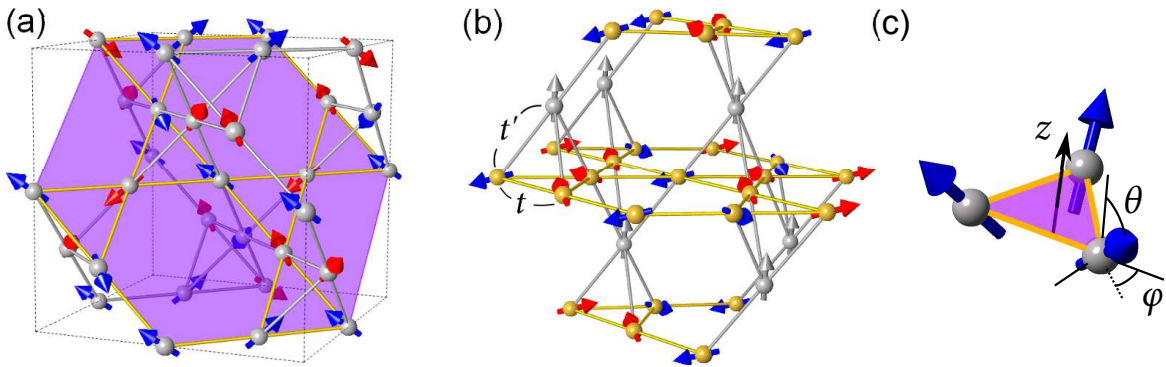}
\end{center}
\caption{(Color online)
(a) Schematic picture of the pyrochlore lattice with localized Ising spins pointing along the local $[111]$ axes.
The kagome plane parallel to the $\langle111\rangle$ plane is shown by the purple shaded plane.
The spin configuration in the plane represents an example of the one-in two-out or two-in one-out states.
(b) Pyrochlore lattice as an alternating stacking of the triangular and kagome lattices. 
The spin configuration gives an example of the kagome-ice states. 
(c) Schematic picture of the localized Ising moments in a triangle on the kagome plane.
See the text for details.
}
\label{fig:model}
\end{figure}

One of such interesting phenomena proposed recently is the kagome-ice insulator with quantum anomalous Hall response~\cite{Ishizuka2013,Chern2012}. 
The kagome-ice insulator appears in the DE model on a kagome lattice with localized Ising moments that correspond to the $\langle111\rangle$ kagome plane of the spin ice [see Fig.~\ref{fig:model}(a)].
In this model, each Ising spin takes either of the two states, ``in" or ``out".
When the Ising spins interact with each other only by the ferromagnetic NN interaction, the ground state remains disordered with macroscopic degeneracy, in which all the triangles obey one-in two-out or two-in one-out spin configurations [see Fig.~\ref{fig:model}(a)].
An external magnetic field perpendicular to the plane partially lifts the degeneracy by enforcing the upward (downward) triangles to be two-in one-out (one-in two-out), but the ground state is still disordered [see the kagome planes in Fig.~\ref{fig:model}(b)]. 
This state is called the kagome ice~\cite{Matsuhira2002,Higashinaka2003,Hiroi2003}.
The peculiar spin texture in the kagome ice gives rise to a charge gap in the electronic structure at a commensurate 2/3 electron filling.
This is the kagome-ice insulator. 
With the gap opening, the Hall conductivity, which is induced by the spin scalar chirality, becomes quantized at a nonzero integer value despite the absence of the magnetic order.
The realization of the kagome-ice insulator at finite temperature was shown~\cite{Ishizuka2013}.
The localized state at a defect was also studied~\cite{Chern2012}.

Given the peculiar kagome-ice insulator, its stability against perturbations is of considerable interest. 
Especially, the effects of the three dimensionality (interlayer coupling) and the canting angle of the Ising spins are interesting from the experimental point of view.
In this study, we numerically investigate these two effects on the kagome-ice insulator in the DE models by using the exact diagonalization method.
We show that the kagome-ice insulator is stable for the weak interlayer coupling.
On the other hand, the gap also remains robust against the change of the canting angle. 
We also find that another gap opens at 1/3 filling for the anisotropy axes closer to the collinear case.

\section{Model and Method}
\subsection{Model}
In Sec.~\ref{sec:pyro}, to investigate the stability of the kagome-ice insulator against the three dimensionality, we consider the DE model in the strong coupling limit~\cite{Zener1951,Anderson1955} on an anisotropic pyrochlore lattice in which the kagome layers are weakly coupled to the triangular layers between them [see Fig.~\ref{fig:model}(b)]. 
The Hamiltonian is given by
\begin{eqnarray}
H = - t \sum_{\langle i,j \rangle} ( \tau_{ij} c^\dagger_{i} c_{j} + \text{H.c.} ) - t^\prime \sum_{\left\{ i,j \right\}} ( \tau_{ij} c^\dagger_{i} c_{j} + \text{H.c.} ),
\label{eq:Hpyro}
\end{eqnarray}
where, $c_i$ ($c_i^\dagger$) is the annihilation (creation) operator of an itinerant electron at $i$th site, whose spin index is dropped as the spin is completely aligned parallel to the localized spin $\mathbf{S}_i$ at each site.  
The sum $\langle i,j \rangle$ is taken over the NN intralayer bonds within the kagome lattices, and $\left\{ i,j \right\}$ is over the NN interlayer bonds connecting the triangular and kagome layers [see Fig.~\ref{fig:model}(b)]. 
The transfer integrals for the intralayer and interlayer bonds are $t$ and $t'$, respectively, both of which are modulated by the relative angle of neighboring Ising spins by a factor of $\tau_{ij} = \cos\frac{\theta_i}{2}\cos\frac{\theta_j}{2}+ \sin\frac{\theta_i}{2}\sin\frac{\theta_j}{2}e^{-{\rm i}(\varphi_i-\varphi_j)}$.
The anisotropy axis of the localized spin depends on the sublattice, as in the spin ice.
For the spins on the kagome plane, ${\bf S}_i=(S_i^x,S_i^y,S_i^z)=S(\sin\theta_i\cos\varphi_i,\sin\theta_i\sin\varphi_i,\cos\theta_i)$, where  $(\theta_i, \varphi_i) = (\arccos(\frac13), \frac{2\pi}{3}n_{\rm s} + \frac{\pi}{2}), (\arccos(-\frac13), \frac{2\pi}{3}n_{\rm s} - \frac{\pi}{2})$ for the sublattice $n_{\rm s}=1,2,3$ of the upward triangles [see Fig.~\ref{fig:model}(c)].
On the other hand, the anisotropy axis of the spin on the triangular planes is perpendicular to the plane; ${\bf S}_i = S(0,0,\pm1)$. We take $S=1$ hereafter.

On the other hand, in Sec.~\ref{sec:cant}, we consider the DE model on a kagome lattice,
\begin{eqnarray}
H = - t \sum_{\langle i,j \rangle} ( \tau_{ij} c^\dagger_{i} c_{j} + \text{H.c.} ),
\label{eq:Hkagm}
\end{eqnarray}
where the sum is limited in a single layer of the kagome lattice.
We study the stability of the kagome-ice insulator against the canting angle $\theta$ of the anisotropy axis for the localized spins while fixing $\varphi_i$ [see Fig.~\ref{fig:model}(c)].

In this study, we focus on the case in which the spin configuration for the localized moments satisfies the kagome-ice configuration, i.e., all the upward (downward) triangles in the kagome planes are in the two-in one-out (one-in two-out) configurations and the spins on the triangular lattice is ferromagnetically ordered parallel to the net magnetic moment of the kagome layers [see Fig.~\ref{fig:model}(b)]. 
Hereafter, we set the energy unit $t=1$, the length of Bravais lattice vector $a = 1$, and the Boltzmann constant $k_{\rm B} = 1$.

\subsection{Numerical diagonalization}

We numerically study the problem by taking a simple average (arithmetic mean) over different spin configurations in the kagome-ice manifold.
For the single-layer model in Eq.~(\ref{eq:Hkagm}), each spin configuration is generated by using the loop algorithm~\cite{Rahman1972,Barkema1998}, starting from the $q=0$ two-in one-out order.
On the other hand, for the three-dimensional model in Eq.~(\ref{eq:Hpyro}), we generate the spin configuration by using the loop algorithm, but limiting the path of the loops in each kagome plane; the spins on the triangular layers are fixed so that the spins are ferromagnetically aligned.
We repeated sufficient numbers of loop updates so that the different spin configurations are statistically independent of each other. 
At each value of the interlayer coupling or canting angle, we calculate the electronic state for the itinerant electrons by numerical diagonalization for a given spin configuration, and take a simple average of the results over different spin configurations. 
The calculations were done for $N_k=8^3$ ($N_k=12^2$) sites of $N=4\times 6^3$ ($N=3\times12^2$) supercell for the model in Eq.~(\ref{eq:Hpyro}) [Eq.~(\ref{eq:Hkagm})]; we took the average over 32 (64) different spin configurations for the model in Eq.~(\ref{eq:Hpyro}) [Eq.~(\ref{eq:Hkagm})].
To estimate the size of charge gap, we calculated the energy differrence between the eigenenergies of the lowest-unoccupied and the highest-occupied states among different samples.
The error bars for the gap size is estimated from the corrected sample standard deviation for different series of the spin configuration.

\section{Results}

\subsection{Effect of interlayer coupling in the anisotropic pyrochlore lattice} \label{sec:pyro}

\begin{figure}
\begin{center}
\includegraphics[width=\linewidth]{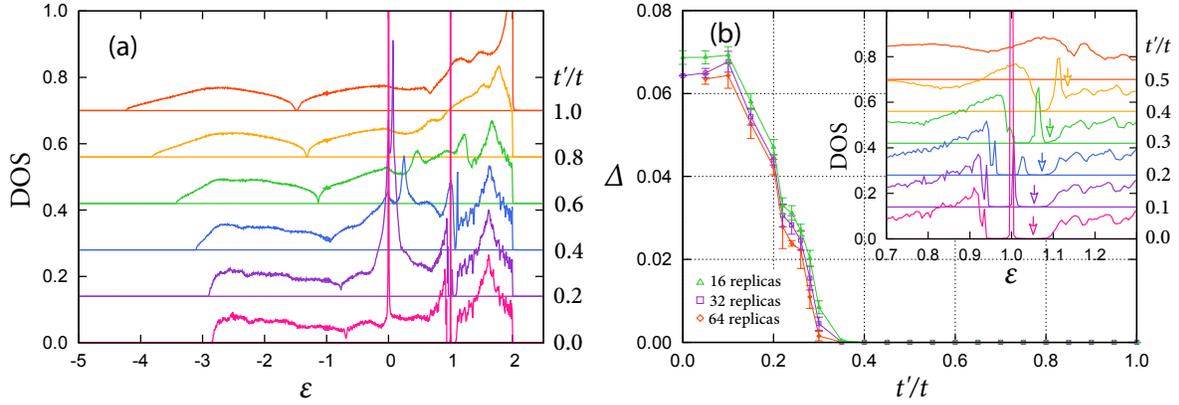}
\end{center}
\caption{(Color online)
(a) The averaged DOS for the kagome-ice state in the anisotropic pyrochlore lattice 
while varying $t^\prime/t$ and (b) the size of energy gaps at $n=2/3$.
The inset in (b) shows the enlarged view of the energy gap at $n=2/3$; the arrows indicate the Fermi level for $n=2/3$.
}
\label{fig:tprime}
\end{figure}

Figure~\ref{fig:tprime}(a) shows the result of the density of states (DOS) calculated while varying $t^\prime/t$.
In the absence of the interlayer coupling $t^\prime/t=0$, two energy gaps are present at the energy around $\varepsilon \sim 1$, which is separated into two by the flat band at $\varepsilon = 1$~\cite{Ishizuka2013,Chern2012}.
In the previous studies on the two-dimensional kagome lattice model, the electron filling just above the flat band corresponds to a commensurate value, $n=\frac1N \sum_i \langle c_i^\dagger c_i \rangle =2/3$.
In the current three-dimensional model at $t'/t=0$, the DOS consists of the contributions from the separated kagome planes and the isolated triangular-lattice sites; the latter corresponds to another flat band at $\varepsilon=0$.
As a consequence, the electron filling at the energy gap just above the flat band at $\varepsilon=1$ is $n=3/4$ instead of $n=2/3$.

The evolution of the size of energy gap at $n=3/4$ with respect to $t^\prime/t$ is shown in Fig.~\ref{fig:tprime}(b).
As increasing $t^\prime$, the energy gap shrinks and vanishes at $t^\prime/t\sim 0.3$.
This result indicates that, though the gap is suppressed by $t^\prime$, it persists up to $t'/t \sim 0.3$.
The enlarged view of the energy gap at $\varepsilon \sim 1$ is also shown in the inset of Fig.~\ref{fig:tprime}(b); the arrows in the inset show Fermi levels for $n=3/4$.
By introducing $t^\prime$, the flat band at $\varepsilon=1$ in the $t^\prime =0$ limit is broadened and shifted toward a higher energy.
At the same time, the gap shrinks and closes at $t'/t\sim 0.3$; for larger $t'/t$, a pseudo-gap like structure remains up to $t'/t\sim 0.5$.
On the other hand, the energy gap below the flat band at $\varepsilon=1$, where $n \sim 0.74$~\cite{note_flatband}, appears to be more robust than the gap at $n=3/4$ and persists up to $t^\prime/t \sim 0.4$.

\subsection{Effect of the canting angle}\label{sec:cant}

\begin{figure}
\begin{center}
\includegraphics[width=\linewidth]{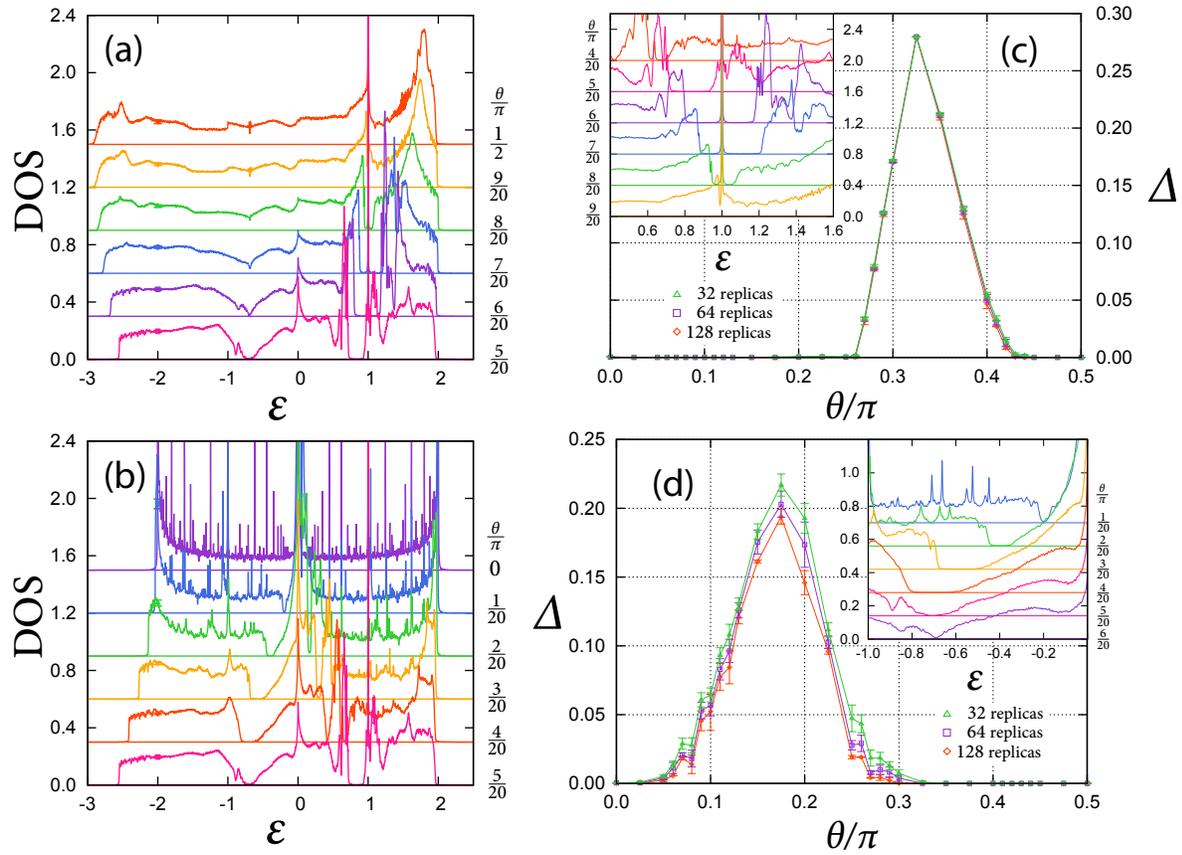}
\end{center}
\caption{(Color online)
(a), (b) The averaged DOS for the kagome-ice states while varying the canting angle $\theta$.
The size of the energy gaps for $n=2/3$ and $1/3$ is shown in (c) and (d), respectively.
The insets in (c) and (d) are the enlarged views of the energy gaps at $n=2/3$ and $1/3$, respectively.
}
\label{fig:angle}
\end{figure}

We next investigate the stability of the kagome-ice insulator with respect to the canting angle of the localized spins by considering the model in Eq.~(\ref{eq:Hkagm}).
Figures~\ref{fig:angle}(a) and \ref{fig:angle}(b) show the averaged DOS for itinerant electrons for different canting angles $\theta$ from collinear ($\theta=0$) to coplanar state ($\theta=\pi/2$).

Figure~\ref{fig:angle}(a) shows the result of the DOS for $\frac5{20}\pi \le \theta \le \frac12 \pi$.
At $\theta=\pi/2$, the anisotropy axes for the localized moments are parallel to the kagome plane and the spins become coplanar; the model with coplanar axes was called the kagome spin ice~\cite{Wills2002}.
The result in this limit shows no gap in the DOS, as shown in Fig.~\ref{fig:angle}(a).

When decreasing $\theta$ from $\pi/2$, the DOS shows an energy gap at $n=2/3$, as found in the previous studies~\cite{Ishizuka2013,Chern2012}.
The detailed evolution of the energy gap and an enlarged view of the DOS are shown in the main panel and the inset of Fig.~\ref{fig:angle}(c), respectively.
Similar to the case in Sec.~\ref{sec:pyro}, two energy gaps appear above and below the flat band at $\varepsilon=1$; the upper gap corresponds to the Fermi level for $n=2/3$ and the lower to $n\sim 0.656$~\cite{note_flatband}.
As seen in the inset of Fig.~\ref{fig:angle}(c), both of the charge gaps appear around $\theta\sim \frac{8}{20}\pi - \frac{9}{20}\pi$.
With further decreasing $\theta$, the energy gap at $n=2/3$ grows rapidly and shows a maxima around $\theta\sim\frac7{20}\pi$, as shown in Fig.~\ref{fig:angle}(c).
The gap starts to shrink for smaller $\theta$ and vanishes at $\theta\sim\frac5{20}\pi$.
In contrast, the gap at $n\sim 0.656$ appears to monotonically grow down to $\theta\sim \frac{5}{20}\pi$ and rapidly decreases below $\theta\sim \frac{5}{20}\pi$; the gap closes at $\theta\sim \frac4{20}\pi$.

On the other hand, in the region with a small $\theta$, another energy gap appears at the Fermi level at $n=1/3$, as shown in Fig.~\ref{fig:angle}(b).
With decreasing $\theta$ from $\theta \sim \frac{5}{20}\pi$, an energy gap appears at the Fermi level at $n=1/3$.
The gap grows with decreasing $\theta$ with maxima around $\theta\sim0.175\pi$.
With further decreasing $\theta$, the energy gap starts to shrink, and vanishes as it approaches $\theta\to 0$.
The dependence of the energy gap on the canting angle $\theta$ is shown in Fig.~\ref{fig:angle}(d).
The result indicates that the energy gap at $n=1/3$ appears in the wide range of $\theta$ for $\frac{1}{20}\pi \lesssim \theta \lesssim \frac{5}{20}\pi$.

At $\theta=0$, the system consists of the 1D loops of up spins and the isolated sites of down spin, as the transfer $
t \tau_{ij}=0$ between up and down spin sites.
Hence, as shown in Fig.~\ref{fig:angle}(b), the DOS consists of the 1D-like electronic DOS from the up-spin loops and the flat band at $\varepsilon=0$ from the down-spin sites; the spikes come from the finite size loops which give discrete energy levels.

Briefly summarizing, the evolution of the electronic DOS with respect to the canting angle $\theta$ shown here indicates that the kagome-ice insulating state at $n=2/3$ is stable in a wide range of $\theta$ for $\frac{5}{20}\pi \lesssim \theta\lesssim \frac9{20}\pi$.
Furthermore, a different charge gap appears at $n=1/3$ for $\frac1{20}\pi \lesssim \theta \lesssim \frac{5}{20}\pi$.
These results indicate that the peculiar charge gap formation induced by a correlated spin texture without magnetic ordering takes place widely in the kagome-ice type models.

\section{Summary}

To summarize, we studied the stability of the quantum Hall kagome-ice insulator in the double-exchange model against the three dimensionality and the canting angle of the localized spins.
In the fore half, we studied the effect of the interlayer coupling by considering an anisotropic pyrochlore lattice model.
We showed that the kagome-ice insulator remains stable up to $t^\prime/t \sim 0.3$, where $t$ ($t'$) is the intra(inter)layer hopping.
In the later half, we studied the stability of the kagome-ice insulator against the canting angle of the localized Ising spins.
We showed that the kagome-ice insulator at $n=2/3$ is stable in the wide range of the canting angle, $ \frac{5}{20}\pi \lesssim \theta \lesssim \frac{9}{20}\pi$, which includes the spin ice case, $\theta=\arccos(1/3)\sim0.392\pi$.
Furthermore, we showed that an energy gap also appears at the Fermi level for $n=1/3$ for $\frac{1}{20}\pi \lesssim \theta \lesssim \frac{5}{20}\pi$.
The results indicate that the quantum Hall response will be seen in a wide range of parameters in the variants of the kagome-ice double-exchange models.

\section*{Acknowledgement}

The authors are grateful to C. D. Batista and Y. Tokura for helpful comments.
H.I. is supported by Grant-in-Aid for JSPS Fellows.
This research was supported by KAKENHI (No. 21340090, 22540372, and 24340076), the Strategic Programs for Innovative Research (SPIRE), MEXT, and the Computational Materials Science Initiative (CMSI), Japan.

\end{document}